\newcommand{\kms}{\ifmmode{\rm km\thinspace s^{-1}}\else km\thinspace s$^{-1}$\fi}

\documentclass{IAU-TrA}
\usepackage{graphicx}

\title[RADIAL VELOCITIES]     
{}

\author[DIVISION~IX / COMMISSION 30]   
{}

\pubyear{2009}
\volume{Volume XXVIIA}
\pagerange{001--008}
\jname{Transactions IAU, Volume XXVIIA}
\editors{Karel A. van der Hucht, ed.}
\begin{document}

\maketitle

{\bf

\large
\begin{tabbing}
\hspace*{65mm}       \=                                              \kill
COMMISSION~30        \> RADIAL VELOCITIES                            \\[0.5ex]
                     \> {\small\it VITESSES RADIALES}                 \\
\end{tabbing}

\normalsize

\begin{tabbing}
\hspace*{65mm}       \=                                              \kill
PRESIDENT            \> Stephane Udry                                \\
VICE-PRESIDENT       \> Guillermo Torres                             \\
PAST PRESIDENT       \> Birgitta Nordstr\"om                         \\
ORGANIZING COMMITTEE \> Francis C. Fekel,                            \\ 
                     \> Kenneth C. Freeman,                          \\
                     \> Elena V. Glushkova,                          \\
                     \> Geoffrey W. Marcy,                           \\
                     \> Birgitta Nordstr\"om,                        \\
                     \> Robert D. Mathieu,                           \\
                     \> Dimitri Pourbaix,                            \\   
                     \> Catherine Turon,                             \\   
                     \> Tomaz Zwitter                                \\   
\end{tabbing}

\noindent
COMMISSION~30 WORKING GROUPS
\smallskip

\begin{tabbing}
\hspace*{65mm}      \=                                               \kill
Div.~IX / Commission~30 WG \>  Radial-Velocity Standard Stars     \\
Div.~IX / Commission~30 WG \>  Stellar Radial Velocity Bibliography     \\
Div.~IX / Commission~30 WG \>  Catalogue of Orbital Elements of  \\
                           \>  ~~Spectroscopic Binaries \\
\end{tabbing}

\bigskip

\noindent
TRIENNIAL REPORT 2006--2009
}

\firstsection 

\section{Introduction}

This three-year period has seen considerable activity in the
Commission, with a wide range of applications of radial velocities as
well as a significant push toward higher precision. The latter has
been driven in large part by the exciting research on extrasolar
planets. This field is now on the verge of detecting Earth-mass bodies
around nearby stars, as demonstrated by recent work summarized below,
and radial velocities continue to play a central role.

This is not to say that classical applications of RVs have lagged
behind. On the contrary, this triennium has seen the release of
several very large data sets of stellar radial velocities (Galactic
and extragalactic) that are sure to have a significant impact on a
number of fields for years to come. The era of mass-producing radial
velocities has arrived. Examples include the Geneva-Copenhagen Survey,
the Sloan Digital Sky Survey, and RAVE, and are described below.

Due to circumstances beyond our control, the report of Commission 30
for the previous (2003--2006) triennium did not appear in the printed
version of the Transactions of the IAU, although it did appear in the
electronic version. For progress during the previous period, the
reader is therefore encouraged to consult the latter, which is
available from the Commission web site.

\section {Radial velocities and exoplanets}

\noindent By G. Torres and J. Johnson
\vskip 9pt

\subsection{Toward Earth-mass planets}

Detections of Jupiter-mass exoplanets by the radial-velocity method
relying on measurements with precisions of a few m~s$^{-1}$ are now
quite routine. This technique has provided by far the majority of the
more than 300 planet discoveries to date. The persistence of
astronomers and the increasing precision of their instruments has led
to larger and larger numbers of multi-planet systems being found. One
example is the interesting case of $\mu$~Arae (\cite{Pepe:07b}), with
\textit{four} planets, one of which is as small as 10.5~M$_{\rm
Earth}$. The host star also presents the signature of $p$-mode
oscillations seen clearly in the radial velocities. The record-holder
for the most planets is the star 55~Cnc, which is orbited by no less
than \textit{five} planets (\cite{Fischer:08}), of which the smallest
has a minimum mass of 10.8~M$_{\rm Earth}$. Exciting discoveries
during this period made possible by the high precision and stability
of the HARPS instrument on the ESO 3.6-m telescope at La Silla (Chile)
include the system Gls~581 (at only 6.3 pc), attended by at least
three planets. In addition to the previously known Neptune-mass body
orbiting the star with a period of 5.3 days, two other low mass
planets were found by \cite{Udry:07a} with minimum masses of only
5~M$_{\rm Earth}$ (period 12.9 days) and 7.7~M$_{\rm Earth}$, the
latter being near the outer edge of the habitable zone of the M3V
parent star (period 83.6 days). The three-planet orbital solution for
this case has an rms residual of only 1.2 m~s$^{-1}$. Another system
with three low-mass planets was announced by \cite{Mayor:08a} around
the nearby (13 pc) metal-poor K2V star HD~40307. The planets weighed
in at 4.2, 6.9, and 9.2~M$_{\rm Earth}$, and the three-planet
Keplerian orbital fit gave impressive residuals of just 0.85
m~s$^{-1}$. HARPS has demonstrated that this sort of velocity
precision is achievable for ``quiet'' stars that present a low level
of ``jitter'' in their radial velocities due to astrophysical
phenomena such as $p$-mode oscillations, granulation, or chromospheric
activity. Indications are that Neptune-mass or smaller planets are
more common around solar-type (F--K) stars than previously thought
(see, e.g., \cite{Mayor:08b}).

\subsection{Retired A stars and their planets}

Most Doppler searches for planets have concentrated on main sequence
stars of spectral types F or later, because the velocity precision for
earlier type stars is seriously compromised by line broadening induced
by rapid rotation, as well as the overall fewer number of spectral
lines available. This difficulty in studying higher-mass stars
introduces a bias in our understanding of planets, but it can be
overcome by looking at such stars after they have left the main
sequence. This is precisely the approach of an ongoing project to
investigate the relationship between stellar mass and planet formation
by using the HIRES instrument on the Keck 10-m telescope to search for
planets in a sample of 240 intermediate-mass subgiants ($1.3 <
M_*/M_\odot < 2.2$). Subgiants have lower surface temperatures and
rotational velocities than their main-sequence progenitors, making
them ideal proxies for A- and F-type stars in Doppler studies. From a
smaller sample of subgiants observed previously by \cite{Johnson:07a}
at Lick Observatory for 4 years with a typical velocity precision of 4
m~s$^{-1}$, a strong correlation was detected between stellar mass and
planet occurrence, with a detection rate of 9\% within 2.5~AU among
the high-mass sample, compared to 4.5\% for Sun-like stars and less
than 2\% for M dwarfs. A paucity of planets within 1 AU of stars with
masses greater than 1.5~M$_\odot$ was found, indicating that stellar
mass also plays a key role in planet migration (\cite{Johnson:07b,
Johnson:08}).  The goal of the expanded Keck survey (with an increased
velocity precision of about 2 m~s$^{-1}$) is to map out the
relationships between stellar mass and exoplanet properties in greater
detail by examining the distribution of planetary minimum masses,
eccentricities, semimajor axes, and the rate of multiplicity around
evolved A stars. If the 9\% occurrence rate is confirmed, some 20--30
new planets should be found in the sample orbiting some of the most
massive stars so far examined by the Doppler technique.

\subsection{Current status and prospects}

In May of 2008 NASA convened the Exoplanet Forum 2008, a meeting of
experts from the US and other countries in eight different
observational techniques related to exoplanet research. The purpose
was to discuss paths forward for exploring and characterizing planets
around other stars, and to provide specific suggestions for space
missions, technology development, and observing programmes that could
fulfill the recommendations of a previously held meeting of the
Exoplanet Task Force (http://www.nsf.gov/mps/ast/exoptf.jsp).  The
reports resulting from these meetings are intended to provide input
for consideration by various advisory committees in the US, and in
particular by the Astronomy and Astrophysics Decadal Survey that is
currently underway.

Radial velocities was one of the eight techniques considered by the
Exoplanet Forum 2008. The corresponding chapter of the report,
available at
\vskip 8pt
\hskip 40pt {\tt http://exep.jpl.nasa.gov/exep\_exfCommunityReport.cfm}~,
\vskip 8pt
\noindent summarized the progress in the field over the last few
years, which is illustrated by a velocity precision of 1 m~s$^{-1}$ or
slightly better achieved so far, led by the Swiss team using the HARPS
instrument on the ESO 3.6-m telescope, and the California-Carnegie
team using the HIRES instrument on the Keck 10-m telescope. The
factors currently limiting the precision were discussed briefly and
have been described in detail by \cite{Pepe:07a}.  They include various
sources of astrophysical noise (stellar oscillations, granulation,
magnetic cycles, collectively known as ``stellar jitter''), guiding,
the illumination of the spectrograph, and the wavelength
reference. Good progress has been made in each of these areas. For
example, it appears that jitter can be substantially reduced through
longer exposures or binning, to the level of perhaps 10 cm~s$^{-1}$ or
less. A new thorium-argon line list was developed by \cite{Lovis:07}
that significantly improves the velocity precision when using this
source as the wavelength reference. Further improvements in the
velocity precision perhaps reaching a few cm~s$^{-1}$ appear possible
using a dense spectrum of lines generated by a femtosecond-pulsed
laser (``laser comb''), described in more detail below.  The next few
years will tell whether this promise can be realized in practice.

The report of the Exoplanet Forum also described recent progress in
techniques to measure precise velocities in the near infrared (see,
e.g., \cite{Ramsey:08}), which are now approaching the 10 m~s$^{-1}$
level in initial tests. Longer wavelengths potentially provide a
significant advantage for the Doppler detection of very small (even
Earth-mass) planets, since these objects produce a larger signal when
orbiting less massive stars, which emit most of their flux in the near
infrared.

In addition to velocity precision, the report pointed out what is
currently considered by the community to be the greatest challenge for
making progress in the detection of exoplanets by the Doppler
technique: the limited access to telescope time. This has a direct
impact not only on the size of the samples of solar-type stars that
can be studied, but also severely restricts the number of late-type
(faint) stars that can be targeted to search for Earth-mass planets.
The need for exposure times longer than dictated by Poisson statistics
to reduce stellar jitter, as mentioned above, is a further strain on
the limited resources currently available on telescopes equipped with
high-precision spectrographs.

\section{Toward higher radial velocity precision}

\noindent By G. Torres
\vskip 9pt

During this period agreement has been reached for the construction of
an improved copy of the very successful HARPS spectrograph, currently
in operation at the ESO 3.6-m telescope at La Silla, for the northern
hemisphere (HARPS-NEF). This is a high-resolution ($R \approx
120,\!000$) fibre-fed optical spectrograph with broad wavelength
coverage (3780--6910~\AA) designed for high radial velocity
precision. HARPS-NEF is a collaboration between the New Earths
Facility (NEF) scientists of the Harvard Origins of Life Initiative
and the HARPS team at the Geneva Observatory. It is expected to be the
workhorse for follow-up of transiting planet candidates for NASA's
\textit{Kepler} mission, and should be operational perhaps in late
2010.  HARPS-NEF is a cross-dispersed echelle spectrograph that will
benefit not only from updates and improvements over the original HARPS
instrument, but in addition it will be installed on a larger telescope
aperture in the northern hemisphere (the 4.2-m William Herschel
Telescope on La Palma, Canary Islands). It is designed for ultra-high
stability (10--20 cm~s$^{-1}$), and like HARPS it will be placed in a
vacuum chamber with careful temperature control.

One of the key factors that determine the precision of the RVs is the
wavelength reference. Existing technologies in the optical (such as
the Th-Ar technique and iodine gas absorption cell) have already
reached sub-m~s$^{-1}$ precision in some cases, but further
improvements are needed if the Doppler method is to reach cm~s$^{-1}$,
as is needed to detect terrestrial-mass planets. A new technology that
has emerged in the last few years and that holds great promise for
providing a very stable reference is that of laser ``frequency
combs''. As the name suggests, a frequency comb generated from
mode-locked femtosecond-pulsed lasers provides a spectrum of very
narrow emission lines with a constant frequency separation given by
the pulse repetition frequency, typically 1 GHz for this application.
This frequency can be synchronized with an extremely precise reference
such as an atomic clock. For example, using the generally available
Global Positioning System (GPS), the frequencies of comb lines have
long-term fractional stability and accuracy of better than
$10^{-12}$. This is more than enough to measure velocity variations at
a photon-limited precision level of 1 cm~s$^{-1}$ in astronomical
objects (see, e.g., \cite{Murphy:07}). This direct link with GPS as
the reference allows the comparison of measurements not only between
different instruments, but potentially also over long periods of
time. To provide lines with separations that are well matched to the
resolving powers of commonly used echelle spectrographs, a recent
improvement incorporates a Fabry-P\'erot filtering cavity that
increases the comb line spacing to $\sim$40 GHz over a range greater
than 1000~\AA\ (\cite{Li:08}). Prototypes using a titanium-doped
sapphire solid-state laser have been built that provide a reference
centreed around 8500~\AA. In practice, of course, Doppler measurements
are also affected by other instrumental problems, so that the value of
this new technology for highly precise RV measurements is still to be
demonstrated. Tests have been initiated during this triennium. Plans
call also for the installation of a laser comb on the HARPS-NEF
spectrograph described above.  Applications of this technique are not
limited to stars. For example, a direct measurement of the expansion
of the Universe could be made by observing {\it in real time\/} the
evolution of the cosmological redshift of distant objects such as
quasars. Such a measurement would require a precision in determining
Doppler drifts of $\sim$1 cm~s$^{-1}$ per year (see, e.g.,
\cite{Steinmetz:08}), which a laser comb can in principle deliver.

\section{Radial velocities and asteroseismology}

\noindent By G. Torres
\vskip 9pt

The significant increase in the precision of velocity measurements
over the past few years, driven by exoplanet searches, has enabled
important studies of the internal constitution of stars through the
technique of asteroseismology. A number of spectrographs now reach the
precision needed for this type of investigation.  During this period
\cite{Bedding:06} observed the metal-poor subgiant star $\nu$~Ind with
the UCLES instrument on the 3.9-m Anglo-Australian Telescope, and with
the CORALIE spectrograph on the 1.2-m Swiss telescope at ESO. The
precision of those measurements ranged from 5.9 to 9.5 m~s$^{-1}$, and
allowed the authors to place constraints on the stellar parameters
confirming that the star has a low mass and an old age. This was the
first application of asteroseismology to a metal-poor
star. $\alpha$~Cen~A was observed by \cite{Bazot:07} with the HARPS
spectrometer on the ESO 3.6-m telescope, and 34 $p$ modes were
identified in the acoustic oscillation spectrum of the star.
Individual observations had errors well under 1 m~s$^{-1}$. A similar
study by \cite{Mosser:08a} was conducted on Procyon ($\alpha$~CMi)
using the SOPHIE spectrograph on the 1.9-m telescope at the
Haute-Provence Observatory, yielding a precision of about 2
m~s$^{-1}$. The HARPS instrument was used again by \cite{Mosser:08b}
to study the old Galactic disk, low-metallicity star HD~203608. A
total of 15 oscillation modes were identified, and the age of the star
was determined to be $7.25 \pm 0.07$ Gyr.

\section{Radial velocities in Galactic and extragalactic clusters}

\noindent By E.\ V.\ Glushkova, H. Levato, and G. Torres
\vskip 9pt

Searches for spectroscopic binaries in southern open clusters have
continued during this period (e.g., \cite{Gonzalez:06}). These authors
have reported results for the open cluster Blanco\,1.  Forty four
stars previously mentioned in the literature as cluster candidates,
plus an additional 25 stars in a wider region around the cluster were
observed repeatedly during 5 years. Six new spectroscopic binaries
have been detected and their orbits determined. All of them are
single-lined spectroscopic systems with periods ranging from 1.9 to
1572~days.  When considering also all suspected binaries, the
spectroscopic binary frequency in this cluster amounts to
34\%. Additional velocities were measured in this cluster by
\cite{Mermilliod:08a}, who obtained a rather similar binary frequency.

Results from long term radial velocity studies based on the CORAVEL
spectrometers have been presented during this period for the open
clusters NGC 6192, 6208, and 6268 (\cite{Claria:06}), as well as for
NGC 2112, 2204, 2243, 2420, 2506, and 2682 (\cite{Mermilliod:07a}).
These studies were complemented with photometric observations in a
variety of systems, and included membership determination and binary
studies. A number of new spectroscopic binaries were discovered, and
their orbital elements were determined. 

Other individual cluster studies in the Milky Way, which we merely
reference here without giving the details due to space limitations,
include: IC~2361 (\cite{Platais:07}), NGC~2489 (\cite{Piatti:07}),
$\alpha$~Per (\cite{Mermilliod:08b}), the five distant open clusters
Ru~4, Ru~7, Be~25, Be~73 and Be~75 (\cite{Carraro:07}), Tombaugh~2
(\cite{Frinchaboy:08a}), the Orion Nebula cluster (\cite{Furesz:08}),
the most massive Milky way open cluster Westerlund~1
(\cite{Mengel:08}), the Galactic centre star cluster
(\cite{Trippe:08}), and the globular clusters M4 (\cite{Sommariva:08})
and $\omega$~Cen (\cite{DaCosta:08}).

This triennium saw the publication of the final results of the 20-year
efforts of J.-C.\ Mermilliod and colleagues to measure the radial
velocities of giant stars in open clusters for a variety of studies
related to their kinematics, membership, and photometric and
spectroscopic properties. A catalogue of spectroscopic orbits for 156
binaries based on more than 4000 individual velocities was published
by \cite{Mermilliod:07b}, based on measurements from CORAVEL and the
CfA Digital Speedometers. Orbital periods range from 41 days to more
than 40 years, and eccentricities are as high as $e = 0.81$. Another
133 spectroscopic binaries were discovered but do not have sufficient
observations and/or time coverage to determine orbital elements. This
material provides a dramatic increase in the body of homogeneous
orbital data available for red-giant spectroscopic binaries in open
clusters, and should form the basis for a comprehensive discussion of
membership, kinematics, and stellar and tidal evolution in the parent
clusters. A companion catalogue (\cite{Mermilliod:08c}) reports mean
radial velocities for 1309 red giants in clusters based on $10,\!517$
individual measurements, and mean radial velocities for 166 open
clusters among which 57 are new. This information, combined with
recent absolute proper motions, will permit a number of investigations
of the galactic distribution and space motions of a large sample of
open clusters.

\cite{Frinchaboy:08b} reported on a survey of the chemical and
dynamical properties of the Milky Way disk as traced by open star
clusters. They used medium-resolution spectroscopy ($R \approx
15,\!000$) with the Hydra multi-object spectrographs on the Cerro
Tololo Inter-American Observatory 4-m and WIYN 3.5-m telescopes to
derive moderately high-precision RVs ($\sigma < 3$ km~s$^{-1}$) for
3436 stars in the fields of 71 open clusters within 3 kpc of the
Sun. Along with the work described in the preceding paragraph, these
represent the largest samples of clusters assembled thus far having
uniformly determined, high-precision radial velocities.

A good deal of activity focused on kinematic analyses of globular
cluster (GC) systems in other galaxies. \cite{Lee:08a} measured radial
velocities for 748 GC candidates in M31, and \cite{Lee:08b} obtained
radial velocities of 111 objects in the field of M60.
\cite{Konstantopoulos:08} obtained new spectroscopic observations of
the stellar cluster population of region B in the prototype starburst
galaxy M82. \cite{Schuberth:06} presented the first dynamical study
of the GC system of NGC~4636 based on radial velocities for 174
clusters. \cite{Bridges:06} measured radial velocities of 38 GCs in
the Virgo elliptical galaxy M60, and \cite{Bridges:07} obtained new
velocities for 62 GCs in M104.

An interesting problem was discussed by \cite{Abt:08}, pointing to a
possible bias in the RVs of many B-type stars. The author looked at 10
open clusters younger than about 30 million years with sufficient
numbers of measured radial velocities, many of them being measured
with CORAVEL, and found that in each case, the main-sequence B0--B3
stars have larger velocities than earlier- or later-type stars.

\section{Radial velocities for field giants}

\noindent By G. Torres
\vskip 9pt

A programme to measure precise radial velocities for 179 giant stars has
been ongoing at the Lick Observatory, with individual errors of 5--8
m~s$^{-1}$ per measurement (\cite{Hekker:06}). This study presented a
list of 34 stable K giants (with RV standard deviations under 10
m~s$^{-1}$) suitable to serve as reference stars for NASA's Space
Interferometry Mission. A follow-up paper (\cite{Hekker:08}) reported
that 80\% of the stars monitored show velocity variations at a level
greater than 20 m~s$^{-1}$, of which 43 exhibit significant
periodicities. One of the goals was to investigate possible mechanisms
that cause these variations. A complex correlation was found between
the amplitude of the changes and the surface gravity of the star, in
which part of the variation is periodic and uncorrelated with $\log
g$, and another component is random and does correlate with surface
gravity.

\cite{Massarotti:08} reported radial velocities made with the CfA
Digital Speedometers for a sample of 761 giant stars, selected from
the Hipparcos Catalogue to lie within 100 pc. Rotational velocities
and other spectroscopic parameters were determined as well. Orbital
elements were presented for 35 single-lined spectroscopic binaries and
12 double-lined binaries. These systems were used to investigate
stellar rotation in field giants to look for evidence of excess
rotation that could be attributed to planets that were engulfed as the
parent stars expanded.

\section{Galactic structure -- Large surveys}

\noindent By B. Nordstr\"om and G. Torres
\vskip 9pt

\subsection{The Geneva-Copenhagen Survey}

During the previous 3-year period one of the mayor surveys completed
and published is the Geneva-Copenhagen Survey of the Solar
Neighbourhood (\cite{Nordstrom:04}). Unfortunately the full
description of this project and the important new science results that
came out of it did not make it into the printed version of the
Transactions of the IAU for 2003--2006, so we summarize and update
that information here for its significant impact for the study of
Galactic structure.  This survey provided accurate, multi-epoch radial
velocities for a magnitude-complete, all-sky sample of $14,\!000$ F
and G dwarfs down to a brightness limit of $V = 8.5$, and is volume
complete to about 40 pc. The catalogue includes new mean radial
velocities for $13,\!464$ stars with typical mean errors of 0.25~\kms,
based on $63,\!000$ individual observations made mostly with the
CORAVEL photoelectric cross-correlation spectrometers covering both
hemispheres. Studies of this rich data set have found evidence for
dynamical substructures that are probably due to dynamical
perturbations induced by spiral arms and perhaps the Galactic
bar. These ``dynamical streams'' (\cite{Famaey:05}) contain stars of
different ages and metallicities which do not seem to have a common
origin. These features, which dominate the observed $U$,$V$,$W$
diagrams, make the conventional two-Gaussian decomposition of nearby
stars into thin and thick disk members a highly dubious procedure. An
analysis by \cite{Helmi:06} suggests that tidal debris from merged
satellite galaxies may be found even in the solar neighbourhood.

A new release of this large catalogue with updated calibrations as well
as new age and metallicity determinations was published during the
present triennium by \cite{Holmberg:07}, and is available from the CDS
at
\vskip 8pt
\hskip 45pt {\tt http://cdsweb.u-strasbg.fr/cgi-bin/qcat?J/A+A/475/51}~.
\vskip 8pt
\noindent A follow-up paper and catalogue are expected to be available
shortly, containing new kinematic data ($UVW$ velocities) resulting
from a re-analysis using the revised {\it Hipparcos\/} parallaxes
(\cite{vanLeeuwen:07}), and online updates.

\subsection{Sloan Digital Sky Survey}

This period saw the sixth data release of the Sloan Digital Sky Survey
(\cite{Adelman:08}), which now covers an area of 9583 square degrees
on the sky. This release includes nearly 1.1 million spectra of
galaxies, quasars, and stars with sufficient signal to be usable,
along with redshift determinations, as well as effective temperature,
surface gravity, and metallicity determinations for many stars. The
spectra cover the wavelength region 3800--9200~\AA\ at a resolving
power ranging from 1850 to 2200. Velocity precisions range from about
9 km~s$^{-1}$ for A and F stars to about 5 km~s$^{-1}$ for K stars.
The zero point of the velocities is in the process of being calibrated
using spectra from the ELODIE spectrograph.  These data are a valuable
resource for a variety of investigations related to Galactic structure
and the evolution and history of the Milky Way.

\subsection{RAVE}

The second data release of the Radial Velocity Experiment (RAVE) was
published during this triennium (\cite{Zwitter:08}). This is an
ambitious spectroscopic survey to measure radial velocities as well as
stellar atmosphere parameters (effective temperature, metallicity,
surface gravity, rotational velocity) of up to one million stars using
the 6dF multi-object spectrograph on the 1.2-m UK Schmidt telescope of
the Anglo-Australian Observatory. The RAVE programme started in 2003,
obtaining medium resolution spectra (median $R = 7500$) in the Ca~II
triplet region (8410--8795~\AA) for southern hemisphere stars drawn
from the Tycho-2 and SuperCOSMOS catalogues, in the magnitude range $9 <
I < 12$. Following the first data release, the current release doubles
the sample of published radial velocities, now reaching $51,\!829$
measurements for $49,\!327$ individual stars observed between 2003 and
2005. Comparison with external data sets indicates that the new data
collected since April 2004 show a standard deviation of 1.3
km~s$^{-1}$, about twice as good as for the first data release. For
the first time, this data release contains values of stellar
parameters from $22,\!407$ spectra of $21,\!121$ individual stars. The
data release includes proper motions from the STARNET 2.0, Tycho-2,
and UCAC2 catalogues, and photometric measurements from Tycho-2, USNO-B,
DENIS, and 2MASS. The data can be accessed via the RAVE web site at
\vskip 8pt
\hskip 105pt {\tt http://www.rave-survey.org}~.
\vskip 8pt
\noindent Scientific uses of these data include the identification and
study of the current structure of the Galaxy and of remnants of its
formation, recent accretion events, as well as the discovery of
individual peculiar objects and spectroscopic binary stars. For
example, kinematic information derived from the RAVE data set has been
used by \cite{Smith:07} to constrain the Galactic escape velocity at
the solar radius to $V_{\rm esc} = 536^{+58}_{-44}$ km~s$^{-1}$ (90\%
confidence).

\section{Working groups}

Below are the progress reports of the three active working groups of
Commission 30. Their efforts are focused on providing a service to the
astronomical community at large through the compilation of a variety
of information related to radial velocities.

\subsection{WG on radial velocity standard stars}

\noindent By S. Udry
\vskip 9pt

Large radial-velocity surveys are being conducted to search for
extrasolar planets around different types of stars, including A to M
dwarfs, and G--K giants (e.g., \cite{Udry:07b}).  Although not aiming at
establishing a set of radial-velocity standard stars, the non-variable
stars in these programmes, followed over a long period of time,
provide ideal candidates for our list of standards. They will moreover
broaden the domain of stellar properties covered (brightness and
spectral type). At this point, the results of most of those programmes
are still not publicly available and we must still wait a bit in order
to fine-tune and enlarge the list presently available at 
\vskip 8pt
\hskip 70pt {\tt http://obswww.unige.ch/$\sim$udry/std/std.html}~.
\vskip 8pt 
\noindent In addition to the by-product aspect of planet search
programmes, a targeted observational effort, dedicated to the definition
of a large sample of RV standards for GAIA, is being pursued with
several instruments (CORALIE, SOPHIE, etc). It will provide in a few
years a list of several thousand suitable standards spread over the
entire sky (\cite{Crifo:07}).

For all of the efforts above, work remains to be done to combine the
data from the different instruments into a common RV system, for
example through the observation of minor planets in the solar system
(\cite{Zwitter:07}). This has still to be done for most of the planet
search programmes, but is already included in the GAIA effort.

\vspace{3mm}

\subsection{WG on stellar radial velocity bibliography}

\noindent By H. Levato
\vskip 9pt

During the 2006--2009 triennium, the WG searched for the papers with
measurements of radial velocities of stars in 33 journals. As of
December 2007 $113,\!658$ entries have been catalogued. We expect to
finish 2008 with more than $150,\!000$. It is worth mentioning that at
the end of 1996 there were $23,\!358$ entries recorded, so that in 10
years the number of entries in the catalogue has expanded by a factor
of five. During the triennium we have improved the search engine to
search by different parameters.  In the main body of the catalogue we
have included information about the technical characteristics of the
instrumentation used for radial velocity measurements, and comments
about the nature of the objects.  The catalogue can be accessed at
\vskip 8pt
\hskip 45pt {\tt http://www.casleo.gov.ar/catalogue/catalogue.html}~.
\vskip 8pt
\vspace{3mm}

\subsection{WG on the catalogue of orbital elements of spectroscopic
binaries (SB9)}

\noindent By D. Pourbaix
\vskip 9pt

In Manchester, a WG was set up to work on the implementation of the
9th catalogue of orbits of spectroscopic binaries (SB9), superseding
the 8th release of \cite{Batten:89} (SB8).  SB9 exists in electronic
format only.  The web site
\vskip 8pt
\hskip 105pt {\tt http://sb9.astro.ulb.ac.be}
\vskip 8pt
\noindent was officially released during the summer of 2001.  This
site is directly accessible from the Commission 26 web site, from BDB
(in Besan\c{c}on), and from the CDS, among others.

Since the last report, substantial progress has been accomplished, in
particular in the way complex systems can be uploaded together with
their radial velocities.  That is the case, for instance, for triple
stars with the light time effect accounted for and systems with a
pulsating primary.

At the time of this writing SB9 contains 2802 systems (SB8 had 1469)
and 3340 orbits (1469 in SB8).  A total of 563 papers were added since
August 2000, although most of them come from {\it outside} the WG.
Many papers with orbits still await uploading into the
catalogue. According to ADS, the release paper (\cite{Pourbaix:04})
has been cited a total of 58 times since 2005.  This is twice as many
as the old Batten et al.\ catalogue over the same period.

Even though this work has been very well received by the community and
a number of tools have been designed and implemented to make the job
of entering new orbits easier (input file checker, plot generator,
etc.), the WG still suffers from a serious lack of manpower.  Few
colleagues outside the WG spontaneously send their orbits (but they
are usually pleased to send their data when we ask for them).  Any
help (from authors, journal editors, and others) is therefore very
welcome.  Uploading an orbit into SB9 also involves checking for
typos.  In this way we have found several mistakes in published
solutions, which we have corrected.  Sending orbits to SB9 prior to
publication (e.g., at the proof stage) would therefore be a way to
prevent some mistakes from making it into the literature.

\vspace{3mm}
 
{\hfill Guillermo Torres}

{\hfill {\it Vice-President of the Commission}}


\begin{thebibliography}{}

\bibitem[Abt(2008)]{Abt:08}
 Abt, H. 2008, \textit{PASP}, 120, 715

\bibitem[Adelman-McCarthy et al.\ 2008]{Adelman:08}
 Adelman-McCarthy, J.\ K., Ag\"ueros, M.\ A., Allam, S.\ S. et al.\
 2008, \textit{ApJS}, 175, 297

\bibitem[Batten et al.\ (1989)]{Batten:89}
 Batten, A.\ H., Fletcher, J.\ M., \& MacCarthy, D.\ G. 1989, Eighth
 catalogue of the orbital elements of spectroscopic binary systems,
 \textit{Publ.\ Dom.\ Astr.\ Obs.}, 17, 1

\bibitem[Bazot et al.\ (2007)]{Bazot:07}
 Bazot, M., Bouchy, F., Kjeldsen, J., Charpinet, S., Laymand, M., \&
 Vauclair, S. 2007, \textit{A\&A}, 470, 295

\bibitem[Bedding et al.\ (2006)]{Bedding:06}
 Bedding, T.\ R., Butler, R.\ P., Carrier, F.\ et al.\ 2006,
 \textit{ApJ}, 647, 558

\bibitem[Bridges et al.\ (2006)]{Bridges:06}
 Bridges, T., Gebhardt, K., Sharples, R.\ et al.\ 2006,
 \textit{MNRAS}, 373, 157

\bibitem[Bridges et al.\ (2007)]{Bridges:07}
 Bridges, T., Rhode, K.\ L., Zepf, S.\ E., \& Freeman, K.\ C. 2007,
 \textit{ApJ}, 658, 980

\bibitem[Carraro et al.\ 2007]{Carraro:07}
 Carraro, G., Geisler, D., Villanova, S.\ et al.\ 2007, \textit{A\&A},
 476, 217

\bibitem[Clari\'a et al.\ 2006]{Claria:06}
 Claria, J.\ J., Mermilliod, J.-C., Piatti, A.\ E., \& Parisi, M.\
 C. 2006, \textit{A\&A}, 453, 91

\bibitem[Crifo et al.\ 2007]{Crifo:07}
 Crifo, F., Jasniewica, G., Soubiran, C.\ et al.\ 2007, in
 \textit{Towards a new set of radial velocity standards for GAIA},
 eds.\ J.\ Bouvier, A.\ Chalabaev, \& C.\ Charbonnel, Proceedings of
 the Annual meeting of the French Society of Astronomy and
 Astrophysics, Grenoble (France), p.\ 459

\bibitem[Da Costa \& Matthew 2008]{DaCosta:08}
 Da Costa, G.\ S., \& Matthew, C.\ G. 2008, \textit{AJ}, 136, 506

\bibitem[Famaey et al.\ 2005]{Famaey:05}
 Famaey, B., Jorissen, A., Luri, X.\ et al.\ 2005, \textit{A\&A}, 430,
 165

\bibitem[Fischer et al.\ 2008]{Fischer:08}
 Fischer, D.\ A., Marcy, G.\ W., Butler, R.\ P.\ et al.\ 2008,
 \textit{ApJ}, 675, 790

\bibitem[Frinchaboy et al.\ 2008]{Frinchaboy:08a}
 Frinchaboy, P.\ M., Marino, A.\ F., \& Villanova, S.\ et al.\ 2008,
 \textit{MNRAS} (in press), arXiv:0809.2559

\bibitem[Frinchaboy \& Majewski (2008)]{Frinchaboy:08b}
 Frinchaboy, P.\ M., \& Majewski, S.\ R. 2008, \textit{AJ}, 136, 118

\bibitem[F\"ur\'esz et al.\ 2008]{Furesz:08}
 F\"ur\'esz, G., Hartmann, L.\ W., Megeath, S.\ T. et al.\ 2008,
 \textit{ApJ}, 676, 1109

\bibitem[Gonz\'alez \& Levato 2006]{Gonzalez:06}
 Gonz\'alez, J.\ F., \& Levato, H. 2006, \textit{RMxAA}, 26, 171

\bibitem[Hekker et al.\ 2006]{Hekker:06}
 Hekker, S., Reffert, S., Quirrenbach, A., Mitchell, D.\ S., Fischer,
 D.\ A., Marcy, G.\ W., \& Butler, R.\ P. 2006, \textit{A\&A}, 454,
 943

\bibitem[Hekker et al.\ 2008]{Hekker:08}
 Hekker, S., Snellen, I.\ A.\ G., Aerts, C., Quirrenbach, A., Reffert,
 S., \& Mitchell, D.\ S. 2008, \textit{A\&A}, 480, 215

\bibitem[Helmi et al.\ (2006)]{Helmi:06}
 Helmi, A., Navarro, J.\ F., Nordstr\"om, B.\ et al.\ 2006,
 \textit{MNRAS} 365, 1309

\bibitem[Holmberg et al.\ (2007)]{Holmberg:07}
 Holmberg, J., Nordstr\"om, B., \& Andersen, J. 2007, \textit{A\&A},
 475, 519

\bibitem[Johnson et al.\ (2007a)]{Johnson:07a} 
 Johnson, J.\ A.\ et al.\ 2007a, \textit{ApJ}, 670, 833

\bibitem[Johnson et al.\ 2008]{Johnson:08} 
 Johnson, J.\ A., Marcy, G.\ W., Fischer, D.\ A.\ et al.\ 2008,
\textit{ApJ}, 675, 784

\bibitem[Johnson et al.\ 2007b]{Johnson:07b} 
 Johnson, J.\ A., Butler, R.\ P., Marcy, G.\ W., Fischer, D.\ A.,
Vogt, S.\ S., Wright, J.\ T., \& Peek, K.\ M.\ G. 2007b, \textit{ApJ},
665, 785

\bibitem[Karchenko et al.\ 2007]{Karchenko:07}
 Karchenko, N.\ V., Scholz, R.-D., Piskunov, A.\ E.\ et al.\ 2007,
 \textit{AN}, 328, 889

\bibitem[Konstantopoulos et al.\ (2008)]{Konstantopoulos:08}
 Konstantopoulos, I.\ S., Bastian, N., Smith, L.\ J.\ et al.\ 2008,
 \textit{ApJ}, 674, 846

\bibitem[Lee et al.\ (2008a)]{Lee:08a}
 Lee, M.\ G., Hwang, Ho S., Kim, S.\ Ch. et al.\ 2008a. \textit{ApJ},
 674, 886

\bibitem[Lee et al.\ (2008b)]{Lee:08b}
 Lee, M.\ G., Hwang, Ho S., Park, H.\ S.\ et al.\ 2008b, \textit{ApJ},
 674, 857

\bibitem[Li et al.\ 2008]{Li:08}
 Li, Ch.-H., Benedick, A.\ J., Fendel, P.\ et al.\ 2008,
\textit{Nature}, 452, 610

\bibitem[Lovis \& Pepe (2007)]{Lovis:07}
 Lovis, C., \& Pepe, F. 2007, \textit{A\&A}, 468, 1115

\bibitem[Massarotti et al.\ (2008)]{Massarotti:08}
 Massarotti, A., Latham, D.\ W., Stefanik, R.\ P., \& Fogel, J. 2008,
 \textit{AJ}, 135, 209

\bibitem[Mayor \& Udry 2008]{Mayor:08b}
 Mayor, M., \& Udry, S. 2008, \textit{Phys.\ Scr.}, 130 (in press)

\bibitem[Mayor et al.\ (2008)]{Mayor:08a} 
 Mayor, M., Udry, S., Lovis, C.\ et al.\ 2008, \textit{A\&A} (in
press), arXiv:0806.4587

\bibitem[Mengel \& Tacconi-Garman 2008]{Mengel:08}
 Mengel, S., \& Tacconi-Garman, L.\ E.\ 2008, in \textit{Young massive
 star clusters - Initial conditions and environments}, Granada, Spain
 (in press), arXiv:0803.4471

\bibitem[Mermilliod et al.\ (2008a)]{Mermilliod:08a}
 Mermilliod, J.-C., Platais, I., James, D.\ J., Grenon, M., \&
 Cargile, P.\ A. 2008a, \textit{A\&A}, 485, 95

\bibitem[Mermilliod \& Mayor 2007]{Mermilliod:07a}
 Mermilliod, J.-C., \& Mayor, M. 2007, \textit{A\&A}, 470, 919

\bibitem[Mermilliod et al.\ (2007)]{Mermilliod:07b}
 Mermilliod, J.-C., Andersen, J., Latham, D.\ W., \& Mayor, M. 2007,
 \textit{A\&A}, 473, 829

\bibitem[Mermilliod et al.\ 2008b]{Mermilliod:08b}
 Mermilliod, J.-C., Queloz, D., \& Mayor, M. 2008b, \textit{A\&A},
 488, 409

\bibitem[Mermilliod et al.\ 2008c]{Mermilliod:08c}
 Mermilliod, J.-C., Mayor, M., \& Udry, S. 2008c, \textit{A\&A}, 485,
 303

\bibitem[Mosser et al.\ (2008a)]{Mosser:08a}
 Mosser, B., Bouchy, F., Marti\'c, M.\ et al.\ 2008a, \textit{A\&A},
 478, 197

\bibitem[Mosser et al.\ (2008b)]{Mosser:08b}
 Mosser, B., Deheuvels, S., Michel, E.\ et al.\ 2008b, \textit{A\&A},
 488, 635

\bibitem[Murphy et al.\ 2007]{Murphy:07}
 Murphy, M.\ T., Udem, Th., Holzwarth, R.\ et al.\ 2007,
\textit{MNRAS}, 380, 839

\bibitem[Nordstr\"om et al.\ 2004]{Nordstrom:04}
 Nordstr\"om, B., Mayor, M., Andersen, J.\ et al.\ 2004,
 \textit{A\&A}, 418, 989

\bibitem[Pepe et al.\ 2007]{Pepe:07b}
 Pepe, F., Correia, A.\ C.\ M., Mayor, M.\ et al.\ 2007,
 \textit{A\&A}, 462, 769

\bibitem[Pepe \& Lovis (2007)]{Pepe:07a}
 Pepe, F.\ A., \& Lovis, C. 2007, in \textit{Physics of Planetary
Systems, Nobel Symposium 135}, in press

\bibitem[Piatti et al.\ 2007]{Piatti:07}
 Piatti, A., Clari\'a, J.\ J., Mermilliod, J.-C., Parisi, M.\ C., \&
 Ahumada, A.\ V. 2007, \textit{MNRAS}, 377, 1737

\bibitem[Platais et al.\ 2007]{Platais:07}
 Platais, I., Melo, C., Mermilliod, J.-C.\ et al.\ 2007,
 \textit{A\&A}, 461, 509

\bibitem[Pourbaix et al.\ 2004]{Pourbaix:04}
 Pourbaix, D., Tokovinin, A.\ A., Batten, A.\ H.\ et al.\ 2004,
 \textit{A\&A}, 424, 727

\bibitem[Ramsey et al.\ 2008]{Ramsey:08}
 Ramsey, L. W., Barnes, J., Redman, S.\ L., Jones, H.\ R.\ A.,
Wolszczan, A., Bongiorno, S., Engel, L., \& Jenkins, J. 2008,
\textit{PASP}, 120, 887

\bibitem[Schuberth et al.\ (2006)]{Schuberth:06}
 Schuberth, Y., Richtler, T., Dirsch, B.\ et al.\ 2006, \textit{A\&A},
 459, 391

\bibitem[Smith et al.\ (2007)]{Smith:07}
 Smith, M.\ C., Ruchti, G.\ R., Helmi, A.\ et al.\ 2007,
 \textit{MNRAS}, 379, 755

\bibitem[Sommariva et al.\ 2008]{Sommariva:08}
  Sommariva, V., Piotto, G., Rejkuba, M.\ et al.\ 2008, \textit{A\&A}
  (in press), arXiv:0810.1897

\bibitem[Steinmetz et al.\ 2008]{Steinmetz:08}
 Steinmetz, T., Wilken, T., Araujo-Hauck, C.\ et al.\ 2008,
 \textit{Science}, 321, 1335

\bibitem[Trippe et al.\ 2008]{Trippe:08}
 Trippe, S., Gillessen, S., Gerhard, O.\ E.\ et al.\ 2008,
 \textit{A\&A} (in press), arXiv:0810.1040

\bibitem[Udry \& Santos 2007]{Udry:07b}
 Udry, S., \& Santos, N.\ C. 2007, \textit{ARA\&A}, 45, 397

\bibitem[Udry et al.\ (2007)]{Udry:07a}
 Udry, S., Bonfils, X., Delfosse, X.\ et al.\ 2007, \textit{A\&A},
 469, L43

\bibitem[van Leeuwen 2007]{vanLeeuwen:07}
 van Leeuwen, F. 2007, \textit{A\&A}, 474, 653

\bibitem[Zwitter et al.\ 2007]{Zwitter:07}
 Zwitter, T., Mignard, F., \& Crifo, F. 2007, \textit{A\&A}, 462, 795

\bibitem[Zwitter et al.\ 2008]{Zwitter:08}
 Zwitter, T., Siebert, A., Munari, U.\ et al.\ 2008, \textit{AJ}, 136,
 421

\end{thebibliography}
\end{document}